# Developing Products Update-Alert System for e-Commerce Websites Users Using HTML Data and Web Scraping Technique

Ikechukwu Onyenwe  Ebele Onyedinma  Chidinma Nwafor  Obinna Agbata

Department of Computer Science, Nnamdi Azikiwe University, Awka, Nigeria.
ie.onyenwe @unizik.edu.ng, eg.osita @unizik.edu.ng, nwaforchidinmaa@gmail.com
ou.agbata @unizik.edu.ng,

## ABSTRACT

*Websites are regarded as domains of limitless information which anyone and everyone can access. The new trend of technology put us to change the way we are doing our business. The Internet now is fastly becoming a new place for business and the advancement in this technology gave rise to the number of e-commerce websites. This made the lifestyle of marketers/vendors, retailers and consumers (collectively regarded as users in this paper) easy, because it provides easy platforms to sale/order items through the internet. This also requires that the users will have to spend a lot of time and efforts to search for best product deals, products updates and offers on e-commerce websites. They have to filter and compare search results by themselves which takes a lot of time and there are chances of ambiguous results. In this paper, we applied web crawling and scraping methods on an e-commerce website to get HTML data for identifying products updates based on the current time. The HTML data is preprocessed to extract details of the products such as name, price, post date and time, etc. to serve as useful information for users.*

## KEYWORDS

NATURAL LANGUAGE PREPROCESSING (NLP), E-COMMERCE, E-RETAIL, HTML, DATA, Web, Webscrapping.

## 1. INTRODUCTION

The advancement of internet technology has enabled the fast growth of e-commerce websites for marketers/vendors and consumers (collectively regarded as users in this paper). Today, shopping over Internet is becoming a common trend and e-retail sales are growing bigger adding to the total retail sales worldwide. This has made buying and selling of items among users easy but it also demands vigorous search of products from the users as the e-commerce industry continues to expand. The best part of e-Commerce is the expansive way websites price, post, and market products and services. This implies that tracking of these e-commerce best parts,by users are becoming a challenging task.

Web scraping is a natural language processing (NLP) technique that describes the use of a program to extract data from HTML files on the internet which can be stored as textfiles or in databases for further analysis. It is usually used at the preprocessing stage of NLP tasks. Unlike the traditional way of extracting data by copying and pasting, web scraping is automated by using programming languages like python by defining some parameters and retrieving data in a shorter time. The scrapped data offers insight into updates such as prices, new products, market dynamics, prevailing trends, practices, etc. employed by the websites users. Access to these data can provide a competitive advantage to a user in the field to which s/he belongs. Since a huge amount of mixed data is constantly being generated on the web, web scraping is widely recognized as an efficient and powerful technique for collecting big data [4].

Typically this data is in the form of patterned data, particularly lists or tables. Programs that interact with web pages and extract data use sets of commands known as application programming interfaces (APIs). These APIs can be `taught' to extract patterned data from single web pages or from all similar pages across an entire web site. Alternatively, automated interactions with websites can be built into APIs, such that links within a page can be `clicked' and data extracted from subsequent pages. This is particularly useful for extracting data from multiple pages of search results. Furthermore, this interactivity allows users to automate the use of websites' search facilities, extracting data from multiple pages of search results and only requiring users to input search terms rather than having to navigate to and search each website first. One major current use of web scraping is for businesses to track pricing activities of their competitors: pricing can be established across an entire site in relatively short time scales and with minimal manual effort. Various other commercial drivers have caused a large number and variety of web scraping programs to have been developed in recent years. Some of these programs are free, whilst others are purely commercial and charge a one-off or regular subscription fee. Such web scraping software are (1) Visual Web Ripper, (2) Web Content Extractor and (3) WebSundew amongst others [3, 4]. Asides web scrapping software, there are programming languages with in-built powerful web scrapping libraries use for building a custom web data extractor with a specific requirement. The most well-known languages used include R (rvest,Rselenium), Python (BeautifulSoup, Selenium), Ruby, NodeJS and Java. These web scraping tools are equally as useful in the e-commerce websites. There are many web scrapping techniques use for raking data on internet. They are (1) Traditional copy and paste, (2) Text grapping and regular expression, (3) Hypertext Transfer Protocol (HTTP) Programming, (4) Hyper Text Markup Language (HTML) Parsing, (5) Document Object Model (DOM) Parsing, (6) Web Scraping Software, (7) Vertical aggregation platforms, (8) Semantic annotation recognizing, (9) Computer vision web page analyzers[1]. Specifically, they can provide valuable opportunities in the search for products updates by making searches of multiple websites (or web pages of a website) more resource-efficient [3].

There are an overwhelming trillions of products updates made daily on e-commerce websites. Hence, it is important for users to be engaged through smart updates alerts to enable efficient detection of frequent updates and selection of required information instantly on e-commerce websites. This allows users to opt-in to timely products updates and to effectively re-engage them with customized and relevant contents. In this paper, we applied web crawling and scraping methods on e-commerce websites for identification of products updates based on the current time. The scrapping scripts are written using python libraries and web crawling works on HTML tags.

**2. Related Work**

In this section, we present some previous work related to web scraping applications in e-commerce and related fields. Web scraping is also called web data extraction. It is a process, which is used to extract large amount of data from websites and to store the extracted data into the local storage in different formats. Web scraping is used for different purposes such as research, analysis of market and comparison of price, collection of opinion of public in business, jobs advertisements, and collection of contact detail of required business. [3] Used web scrapping to increase transparency and resource efficiency in building and sharing protocols that extract search results and other data from web pages for those looking for grey literature. [2] Proposed an intelligent system to automatically monitor the firms' engagement in e-commerce by analyzing online data retrieved from their corporate websites. This proposed system combines web.

---

[1] https://en.wikipedia.org/wiki/Web scraping

content mining and scraping techniques with learning methods for Big Data. [7] developed a scraper software that is capable of collecting the updated information from the target products hosted in fabulous online e-commerce websites. It is implemented using Scrapy and Django frameworks. The authors claimed that the scraper provides the ability to search a target product in a single consolidated place instead of searching across various websites. [5] used a scrappy and kibana/elastic search interface to crawl and scrape a major online clothes retailer. The authors were able to extract 68 text-based field describing a total of 24,701 clothes to help provide precise estimations of fibres types and color frequencies in less than 24hrs. The extracted data revealed that cotton, polyester, viscose and elastane are the 4 main types of fibres used in the textile industry. [6] used scrapers, an automated programs that mechanically traverse the website and steal the data from websites, in the development of a security mechanism using time and byte entropy analysis. The proposed approach is evaluated by various experiments and the result analysis reveals that the proposed approach is efficient in differentiating price scrapers from human users. [1] adopted web crawling and scraping techniques to collect detailed product information from e-Commerce websites for the purpose of comparison to help users to grab their desired products for the best affordable price.

## 3. Methodology

### 3.1 Data Collection

There are five main steps involved in the process of scraping data from the pages of an e-commerce website. For this experiment, we used *tori.fi*, one of the top sites for online shopping in Finland. The following activities are undertaken: mapping selected web pages, developing web scraping source code and process the scraped data. These activities are illustrated in Figure1.

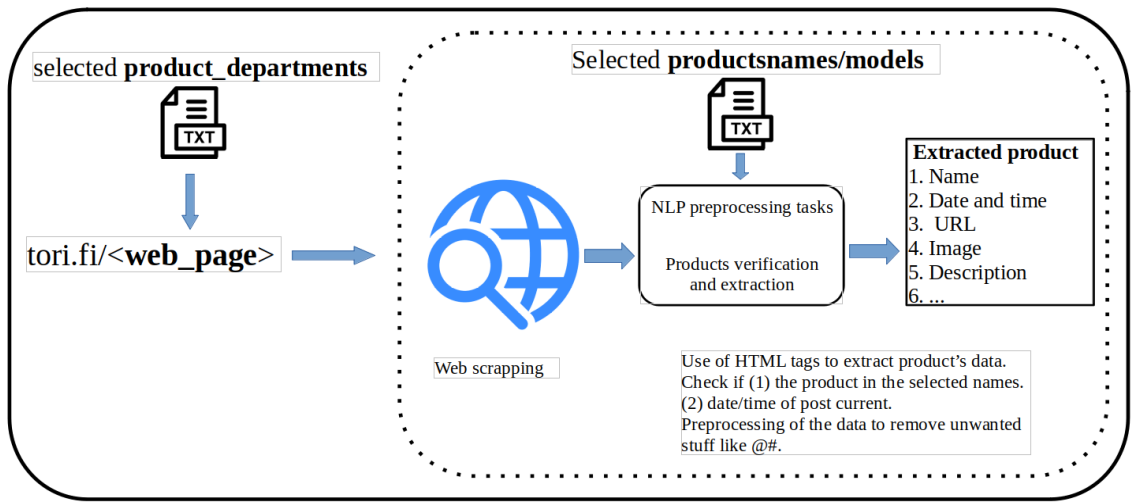

Figure 1: System Architecture of the products update-alert

### 3.2 Mapping Selected Web Pages

E-commerce websites usually categorize their products into different departments such as housing, car, pets, clothes, etc. We created an empty text file called *products_departments* (see figures 1 and 2) and extracted all the departments interested to us into the text file. Also, we created another empty text file called *productnames* (see Figures 1 and 2) and extracted all the products in each departments that are interested to us into it. Furthermore, we understudied the source code of the web pages through a web browser and identified the patterns of HTML tags on a few randomly selected web pages. The identification of the HTML tags are based according to the data attributes to be scrapped. Figure 2 shows an example of some of tags (a, title) identified on some of the target web pages. These tags are added to the attributes of the data to be scrapped.

```
<a tabindex="-1" href="/vaihtoautot/toyota/yaris/84905081" title="Toyota Yaris" aria-label="Toyota Yaris" class="adCard_anchor__2R5Cs block px-2 py-2 m:py-4 m:px-4 l-px-6">

<a tabindex="-1" href="/vaihtoautot/volkswagen/transporter/86101406" title="Volkswagen Transporter" aria-label="Volkswagen Transporter" class="adCard_anchor__2R5Cs block px-2 py-2 m:py-4 m:px-4 l-px-6">
```

Figure 2: Sample source code web page view

**3.3 Developing Web Scrapper**

When we scrape the web data, we write code that sends a request to the server that is hosting the specified web page. The server will return the HTML source code for the web pages we requested. For this purpose, we used Beautiful Soup (BS) for web scrapping in Figure 1. It is a Python library for pulling data out of HTML and XML files. It works with a parser of your choice to provide idiomatic ways of navigating, searching, and modifying the parse tree. It commonly saves developers hours or days of work going through a website for data collection and analysis. Our code is designed to pass data scraping based on the HTML tags (e.g., id, class, a, title, ...) following these steps in sequence iteratively:

1. Request the content of a specific uniform resource locator (URL) from the server. URLs are reformatted web pages of departments listed in the *products_departments*.
2. Download the content that is returned.
3. Identify the attributes which helps to navigate through the tree and find the desired elements of the web page that we want.
4. Verify if products/models are in the *productnames* and if date/time correspond to the current date/time.
5. Extract contents of those elements and analyze.
6. Discard the web page.

**3.4 Process Scrapped Data**

From the results of section 3.3, extracted contents of the verified products are preprocessed to remove HTML tags. A web page is just a text file in HTML format and HTML-formatted text is ultimately just text. To pull out the text from HTML encoding, we used *BeautifulSoup()* function to parse html files. We used different methods of *BeautifulSoup()* to get certain tags, then texts between them are collected. For example, table 1 shows the sample code of our work.

```
...
r = requests.get(hre_)
soup = BeautifulSoup(r.content, "html.parser")
for div in soup._ndAll("div", class ="date-cat-container"):
    for div in div._ndAll("div", class = "date image"):
...
```

Table 1: Sample web scraping code

From Table 1, the code, written in Python, was used to get texts and image URLs of products that certify item 4 of section 3.3. We used *findAll()* to loop through the *soup* contents and find all the *div* tags containing *div* with *date-cat-container* class. Again, we find all the nested *div* with *date_image* and the texts between the $<div> ::: </div>$ and the image URLs are collected by the *get_text()* method.

## 4. Results

As seen in section 3, there are 3 steps we took to scrape, extract and process HTML data from *tori.fi* e-commerce website. Using sections 3.2 and 3.3, uniform resource locators (URLs) of targeted products are identified. Then the details of the products are scraped and processed using method detailed in section 3.4. Figure 3 shows the sample of the results.

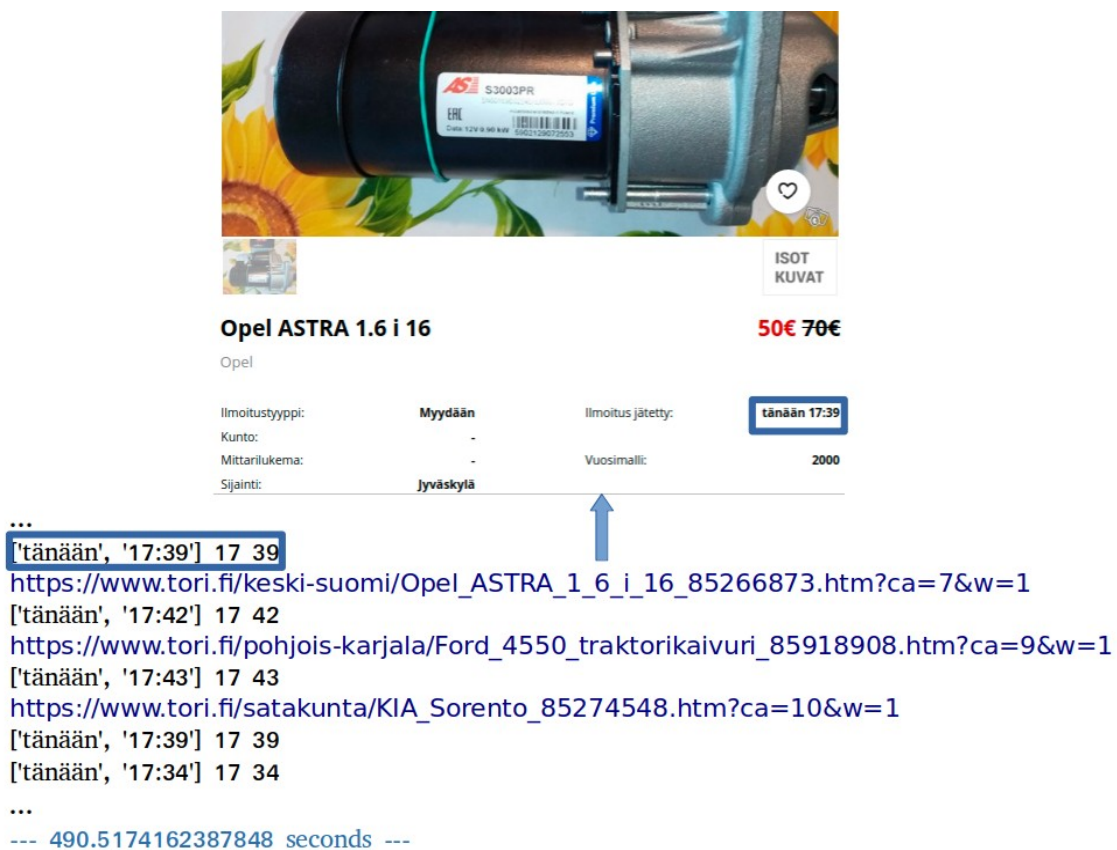

Figure 3: Sample products URLs identified and extracted based on current time. Product's textual details are processed from the scraped HTML data of these web pages.

Observe from the figure, all the URLs are car related (Opel, Ford, KIA, ...) since our *product_departments* and *productnames* in Figure 1 are *autot* in Finnish[2] meaning *car pool* in English and car models for different company cars. *tänään 17:39* in the bold blue rectangle means *today @17:39* time. That is, the product was identified and extracted today at 17:39

---
[2] we used tori.fi, a Finland big e-commerce websites.

Finland time. For those that are not in Finland and wishes to apply this method can adjust the time based on their time zone. For Nigeria, since Finland time is 2 hours ahead, we can add 2 to the current time. We calculated our current time using Python time method *datetime.datetime.now()*. Furthermore, the content of the products as seen on the web page of *Opel ASTRA 1.6i 16* in Figure 3 extracted for user's processed information.

## 5. Conclusion

From the above result presentation, this method/ application for e-commerce websites users can be used to improve products search operations in terms of new/edit products updates information such as sales and costs. It must be said that without the use of the presented method or the likes already stated in the literature, performing products search for tasks such as price comparison process in e-commerce marketplaces would take a long time. Moreover, as prices and products information change very frequently, the obtained information has a very limited time value, and the competitors prices and the middle men sellers (in case of vendors standing in-between companies and consumers) should be analyzed daily in order to take optimal decisions.